\documentclass[preprint2]{aastex}
\usepackage{spr-astr-addons}
\usepackage{graphicx}
\begin{document}
\title{The Wide Field Spectrograph (WiFeS)}

\shortauthors{Dopita et al.}

\author{Michael Dopita , John Hart, Peter McGregor, Patrick Oates \& Gabe Bloxham}
\affil{Research School of Astronomy \& Astrophysics, Australian National University, Cotter Rd., Weston ACT, Australia 2611}

\author{Damien Jones}
\affil{Prime Optics, 17 Crescent Rd, Eumundi, QLD  4562,  Australia}
\email{Michael.Dopita@anu.edu.au}

\begin{abstract}
This paper describes the Wide Field Spectrograph (WiFeS) under construction at the Research School of Astronomy and Astrophysics (RSAA) of the Australian National University (ANU) for the ANU 2.3m telescope at the Siding Spring Observatory. WiFeS is a powerful integral field, double-beam, concentric, image-slicing spectrograph designed to deliver excellent thoughput, wavelength stability, spectrophotometric performance and superb image quality along with wide spectral coverage throughout the $320-950$nm wavelength region.  It provides a $25\times38$~arcsec. field with 0.5~arcsec. sampling along each of twenty five $38\times1$~arcsec slitlets.  The output format is optimized to match the $4096\times4096$~pixel CCD detectors in each of two cameras individually optimized for the blue and the red ends of the spectrum, respectively. A process of ``interleaved nod-and-shuffle" will be applied to permit quantum noise-limited sky subtraction. Using VPH gratings, spectral resolutions  of 3000 and 7000 are provided.  The full spectral range is covered in a single exposure at $R=3000$, and in two exposures in the $R=7000$ mode. The use of transmissive coated optics, VPH gratings and optimized mirror coatings ensures a throughput (including telescope atmosphere and detector) $> 30$\% over a wide spectral range. The concentric image-slicer design ensures an excellent and uniform image quality across the full field. To maximize scientific return, the whole instrument is configured for remote observing, pipeline data reduction, and the accumulation of calibration image libraries.  
\end{abstract}



\section{Introduction}
\label{intro}
An integral field spectrograph is an instrument designed to obtain a spectrum for each of the spatial elements it accepts. In this sense, a classical long-slit spectrograph could also be regarded as an integral-field spectrograph. However, its field shape is rather inconvenient for many practical purposes, being limited to one spatial element in the dispersion direction. The basic problem in designing a grating approach to an integral field spectrograph is therefore to re-format the entrance aperture to allow the acceptance of a convenient 2-D array of spatial elements. 

Broadly speaking, three approaches have been applied to the problem of integral field spectroscopy, which can also be referred to as hyper-spectral imaging. These are, micro-lens array pupil imaging (as in the SAURON spectrograph \citep{Bacon01}, micro-lens arrays coupled to fibre spectrograph feeds \citep{Arribas00,Roth00}, and optical image-slicing \citep{Content00}. 

Each of these have different strengths and weaknesses. The micro-lens array pupil imaging approach allows an instrument of very high throughput, but has a problem of separating the spectra of the individual spatial elements from each other on the detector. This leads to limited spectral coverage, an inefficient use of the detector real estate,  and the data reduction is difficult. 

The fiber-coupled approach as exploited, for example, in the AAOMEGA spectrograph \citep{Saunders04} allows for a free formatting of the spatial elements on the sky, either as a multi-object spectrograph, or as a filled aperture integral field device. The data formatting with this type of device is very convenient, looking very much like a standard long-slit spectrograph, and allowing for easy data reduction. However, the f-ratio of the beam entering the fibers is often large, and any curvature in the fibers will re-distribute the light in an unpredictable way within the effective f-ratio of the fibers themselves. Thus, in order to capture the emergent beam, the f-ratios of the collimator and camera must also be small, which leads to difficult optics design. Variable (and wavelength dependent) fiber throughput and output beam characteristics can lead to a difficult and sometimes unpredictable spectrophotometric calibration and performance. 

The image-slicing approach has the advantage of preserving the input f-ratio, and provides a convenient data format. However, depending on the slicer optics, it may produce a field-dependent image quality. However, this problem can be avoided using the ``concentric" image-slicer concept as first applied to the NIFS instrument located Gemini North \citep{McGregor99,McGregor03}, and now producing exciting science results \citep{McGregor07}.

The success of the Systemic Infrastructure Initiative proposal for the upgrade of the ANU and UNSW telescopes at Siding Spring Observatory (SSO) enabled us to undertake the construction of an entirely new spectroscopic instrument to be mounted at one of the Nasmyth Focii of the 2.3m telescope. It is designed to take maximal advantage of the properties of that telescope while providing a much greater efficiency than the previously available Double Beam Spectrograph (DBS) \cite{Rodgers88}. This paper describes the science drivers, design philosophy, optics and mechanical implementation and the expected on-telescope performance characteristics of this $\mathbf{WI}$de $\mathbf{F}$i$\mathbf{E}$ld $\mathbf{S}$pectrograph (WiFeS).

   \begin{figure*}
   \begin{center}
    \includegraphics[width=11cm]{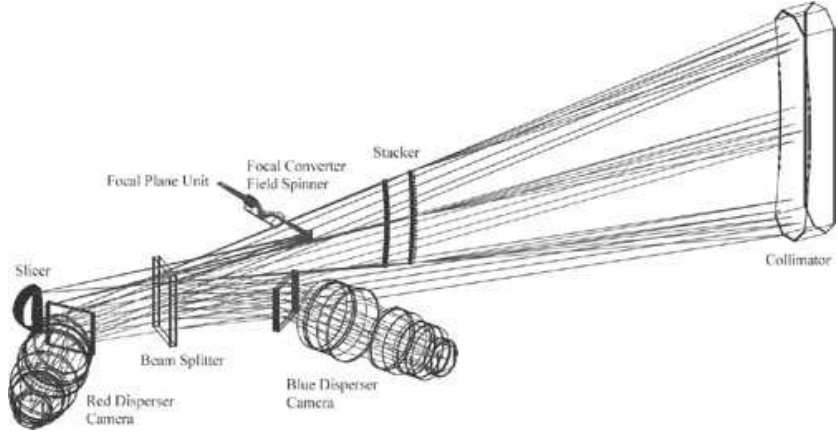}
    \end{center}
   \caption[example] 
   { \label{fig:config} 
  The general optical configuration of the WiFeS spectrograph. Light passes from the field definition
dekker through the de-rotator and f-ratio change optics, and is folded through 90 degrees for packaging
reasons. An image of the entrance slot is projected onto the image slicer which fans each slice 
concentrically to form a set of pupils stacked vertically above one another. The `stacker' refocusses 
the light to form a set of slit images equally spaced on a spherical surface. A slit mask placed here 
eliminates scattered light. The light is then collimated by a spherical collimator, and on its return 
path passes through a dichroic before forming a `blue' and a `red' pupil on VPH gratings. All gratings
are fixed at Littrow and run at the same angle of deviation.}
   \end{figure*}

\section{Science Drivers \& Design Philosophy}
 The 2.3m telescope at Siding Spring Observatory is a relatively modest-sized telescope, with a rather restrictive ($\sim6$ arc min.) unvignetted field of field. The science mission of any new 2.3m instrument should therefore be motivated by the requirement to do exciting and internationally competitive science. This immediately limits the possible instruments to one of two kinds:
\begin{itemize}
\item{Niche instruments designed to do a timely class of science very efficiently, \emph{or}}
\item{An efficient facility-class instrument designed to provide a versatile capability over a whole range of scientific objectives.}
 \end{itemize}
The research and research training mission of the RSAA and of the other user institutions demands the choice of second type of instrument in order to benefit the greatest user community. In particular, the Australian user community is looking for a versatile, stable and efficient instrument which provides not only excellent performance on extended objects, but which can also be used for spectroscopy of single stellar sources with an efficiency appreciably greater than the previously available Double Beam Spectrograph (DBS) \cite{Rodgers88}. 

The science mission that is enabled by such an instrument is very broad. Galactic studies include the internal dynamical studies to investigate mergers, and to study the rotation curves of dwarf galaxies to resolve the core-cusp controversy. The \ion{H}{2} regions in galaxies will be studied to measure abundances and abundance gradients from strong lines, weak lines and recombination lines to resolve the origin of the discrepancies between these three methods. The stellar component of galaxies will be studied to derive spatially resolved spectral energy distributions (SEDs), attenuation by dust, and local star formation rates in starburst galaxies and circum-nuclear star-forming regions. The  gas kinematics and excitation around active nuclei will be studied to provide physical conditions in the ISM of these regions, to quantify the photon and mechanical energy output from the central object. In a few cases it may be possible to estimate the black hole mass.  At high redshifts, we would like to study the dynamics of the extensive Ly$-\alpha$ halos seen round distant radio galaxies, and to study the interstellar or inter-galactic absorption features seen in the GRB sources.

In stellar studies we would like to investigate single stars, especially those hosting planetary systems, and perform studies of the evolution of the early galaxy by observing very metal-poor star candidates identified in surveys, investigating old stellar streams in the halo to investigate the merging history of the Galaxy, to obtain spectroscopy of many stars in a single exposure in Galactic globular clusters, and also to study clusters in the Magellanic Clouds. 

Finally ISM studies include the investigation of the internal structure, excitation and dynamics of planetary nebulae and old nova shells, making abundance analyses of extra-galactic \ion{H}{2} regions as described above, mapping the dynamics and heavy element content of supernova remnants, and studying the outflows seen in Herbig-Haro objects.

These science drivers define the following set of basic science requirements for the spectrograph:
\begin{itemize}
\item{An integral field $\sim 30\times30$ arc sec. or greater.}
\item{Excellent stability of the spectrograph in both the spatial and spectral planes.}
\item{High spectrophotometric veracity ($<1\%$) and excellent sky-subtraction capability.}
\item{The provision of a pipeline data reduction capability and calibration libraries.}
\item{An ability to cover full spectral range with a single exposure at a resolution $R > 2000$.}
\item{The provision of an intermediate resolution mode offering $R \sim 7000$.}
 \end{itemize}
 
The WiFeS spectrograph was designed around the science mission to provide all of these capabilities. Specifically, we have maximized the Felgett Multiplex Advantage; the number of independent spectral and spatial resolution elements. We maximize the detector real estate, compatible with optical constraints and match the spatial resolution to the seeing. The detector scale is 0.5 arc sec per pixel, and the slices are each 1.0 arc sec wide. The Jaquinot Advantage, sometimes known as the luminosity resolution product \citep{Jaquinot60} is also maximized. The luminosity of an instrument is measured by the product of the \'etendu, the solid angle accepted without degradation of the resolution, and the throughput of the instrument. The \'etendu is determined by the nature of the dispersive element. Throughput is maximized by the choice of technology of the coatings and materials of the optical train and in the detector.

Finally, we aim to maximize the science data gathering and evaluation efficiency by providing an instrument with high spectrophotometric stability, simple calibration procedures and operating modes, calibration libraries, efficient feedback of data quality to observer, and batch data reduction to produce standardized data products.

\section{Instrumental Overview}

The WiFeS instrument, scheduled for  completion at the end of 2007 draws upon the design heritage of both the DBS \cite{Rodgers88} and the concentric image-slicer concept of the NIFS instrument for Gemini North \cite{McGregor99,McGregor03}. The instrument is designed to deliver as many simultaneous  spectra as possible, each covering as wide a spectral range as possible. This predicates the use of a  dichroic beamsplitter in front of two gratings and their associated cameras.

The science requirements mandate the use of large-format detectors to provide the required spatial and spectral coverage, but ultimately the maximum field and spectral coverage equally limited by both optical design considerations and mass, and  physical packaging constraints. The detector format for both the ``blue" and the ``red" cameras is $4096\times4096$ with a pixel size of 15$\mu$m square. The need to provide a nod-and-shuffle capability for sky subtraction (which requires half the detector real estate for the object exposures and the other half for the sky reference exposures), together with the need to properly  sample the image (2 pixels) in both the spatial and spectral domains, limits the independent spectral pixels to 8192 and spatial pixels to 1900 per exposure, respectively. 

The optical design aims to critically sample the point spread function delivered by the telescope and the average seeing at Siding Spring, implying a 0.5 arc sec spatial pixel. It also aims to critically sample each resolution element in the spectral domain. Each 1.0 arc sec wide slice is therefore designed to project to 2 pixels on the detector, and the dispersion is chosen to give a resolution which is well-matched to the slit width.
 
Since we wish to provide a relatively simple data reduction pipeline and an instrument with excellent spectrophotometric precision ($<1\%$), the use of fibre-optics were avoided. Instead, a concentric image-slicer design utilising ``long-slit''slitlets was adopted. This concept is shown in Figure \ref{fig:config}.

The requirement for excellent stability of the spectrograph precludes rotation the instrument with the field rotation at the Nasmyth focus. Therefore an Abb\'e-K\"oenig derotator prism is placed just behind the entrance slot mask. The doublet f-ratio conversion lens optically coupled to this avoids an extra air-glass surface.

The requirements for high throughput and good broad-band response predicates the use of transmissive optics where possible, application of optimised optical coating materials,  the use of high-efficiency volume phase holographic (VPH) gratings, and a double-beam operational capability with separate cameras to maximize response in both the ``blue'' and the ``red'' arms and have the capability to collect data from both arms simultaneously.
 
\begin{figure}
   \begin{center}
   \includegraphics[height=5cm]{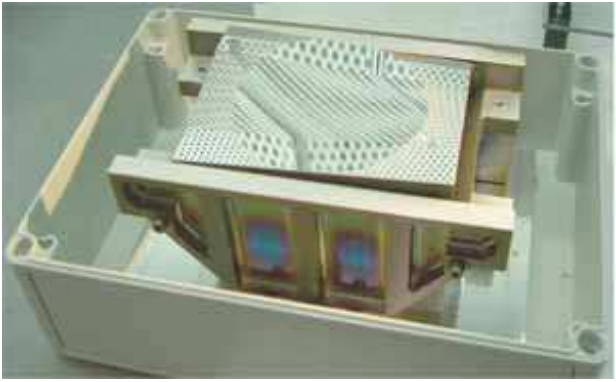}
   \end{center}
   \caption[example] 
   { \label{fig:slicer} 
  The WiFeS image slicer in its final fanned configuration mounted in its mandril which also serves as a mounting block for the slicer in the instrument.}
   \end{figure}

The concentric image-slicer is a macroscopic version of the concept used in the NIFS spectrograph, with the physical width of each slice being 1.75~mm. The image rotator is manufactured as a set of identical 1.75~mm thick half cylindrical plates sandwiched between thicker slabs of glass of the same form.  These are held together in a mandril which applies the lateral pressure and locates all the hemispheres in a vee groove. The multi-element front surface is ground and polished to the correct spherical  figure. The wax is individual slices are fanned out using a pair of adjustment screws at the ends of each one. The vee groove ensures that each slice is rotated about its front face. The fanned slicer is then compressed in the mandril, and coated to provide high reflectance. The final image slicer in its fanned position is illustrated in Figure \ref{fig:slicer}.

The image-slicer forms a set of 25 pupils, one for each slice, stacked vertically above one another on the surface of a sphere centred on the slicer. An achromatic doublet lens and a singlet field lens then forms a row of images of each slitlet on a sphere again centered on the slicer. At this point a slit mask is placed to baffle any stray light in the system. We call this unit the `stacker'. The row of slit images is formed half-way between the slicer and the  collimator. In this configuration, the ray bundle associated with each slitlet strikes the collimator  at normal incidence and the returning collimated beams from all slitlets form a common pupil at the plane of the grating which is  co-planar with the image slicer. By minimizing off-axis angles, this `concentric' approach ensures  excellent and uniform image quality  across the field. 
 
\begin{figure}
   \begin{center}
    \includegraphics[height=7cm]{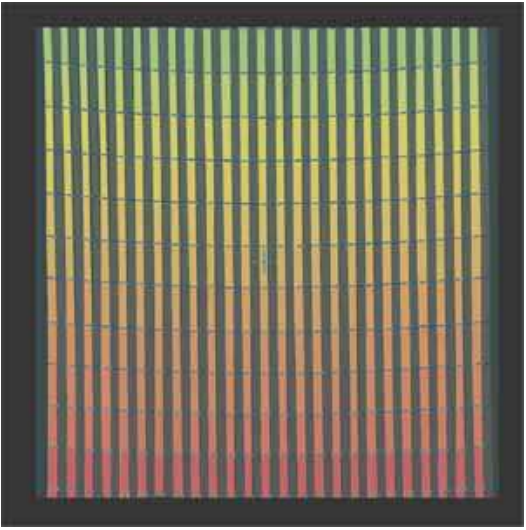}
   \end{center}
   \caption[example] 
   { \label{fig:format} 
 The image format on the detector for the $R_{7000}$ grating. Other gratings give very similar formats. For each image slice the positions on the detector of ten equispaced wavelengths between 5290\AA and 7060\AA\ are shown. Note the spaces between adjacent spectra allow `nod-and-shuffle' sky subtraction to be achieved. The sky spectra are accumulated in these spaces, and sky subtraction is acheived by shifting the image to the left by 80 pixels, and subtracting it from the unshifted image.}
   \end{figure}

The slicer and stacker presents to the WiFeS spectrograph what is in effect a very long-slit format with separated segments corresponding to each slice. Indeed, the effective slit is so long that field angle effects on the grating are appreciable. The fan angle on the image slicer is chosen so that the slitlet image at the stacker is separated by just over its own length from its neighbours. Together these provide an image format on the detector shown in Figure \ref{fig:format}.

This format may look wasteful from the viewpoint of efficient use of the detector real estate, but it does offer a notable advantage for the detection of very faint objects, since it allows for what we have termed ``interleaved nod-and-shuffle" on the CCD. The technique of nod-and-shuffle is well-developed \cite{Cuillandre94,Tinney00,Glazebrook01}, and provides for excellent sky-subtraction because, for any spatial or spectral position, both object and sky are observed virtually at the same time, for the same time, and on the same pixel. This allows for $\sqrt N$ statistics in the removal of unwanted sky signal from the object signal.

The implementation of nod-and-shuffle in WiFeS is to expose for a short period of time, typically 30s. The shutters are closed, and the charge on the CCDs transferred in the $x-$direction by the width of each slitlet ($\Delta x$ pixels) to move the object signals into the inter-slit space.  The telescope is nodded to the reference sky position and a sub-exposure of the same length is taken.  The shutter is again closed, charge shuffled back to its original coordinate, and the telescope returned to  the target position. The whole process is then repeated as many times as necessary. Sky subtraction is achieved by subtracting each part of object signal from the signal located $\Delta x$ pixels from the object signal.  An on-frame bias signal is provided along the chip edge where the charge shuffling causes charge to be repeatedly  lost `overboard'.

Since there remains a risk that the de-rotator axis and the telescope field rotation axis are not precisely aligned, both offset guiding and through entrance slot acquisition cameras are provided (but not shown in Figure \ref{fig:config}. The second camera is fed by a flip mirror immediately in front of, and mechanically coupled to, the image slicer. This enables the observer to re-centre the stellar target or offset guide star precisely on the image slicer between science integrations to avoid any blur resulting from uncorrected field rotation.
 
\section{Optical Considerations} 
\subsection{Concentric Image Slicer Geometry}
The basic geometrical conditions of the IFU and spectrograph are specified as follows. The diameter of the collimated beam in the spectrograph is dictated by the resolving power requirement. The high-resolution gratings operate at a grating angle of $\theta =22^o$, and are required to provide a spectral resolving power of $R = 7000$. The VPH gratings operate in Littrow condition. For a telescope diameter of $D_{\rm tel}$ and an angular slit width of $\delta \phi_{\rm x}$, the required collimator beam diameter, $d_{\rm coll}$, is given by:
 
 \begin{equation}
 d_{\rm coll} = {\frac{D_{\rm tel}\delta \phi_{\rm x} R}{2 \tan\theta}}. \label{eqn1}
 \end{equation}
With $D_{\rm tel}=2300$mm and $\delta \phi_{\rm x} = 1.0$arc~sec., we have $d_{\rm coll} =97$mm.

The angular field size is designated as $\Delta \phi_{\rm x}$ in the spectral direction and $\Delta \phi_{\rm y}$ in the spatial direction. The number of slices determines the aspect ratio of the field. We aim for an aspect ratio $\sim 1.5$ to provide a good format for astronomical observation. With our choice of $N = 25$ slices, a sampling of 2 pixels per slit, and a small clearance margin at the edge of the detector, the number of pixels used in the spatial direction is $n_{\rm spatial} = 4080$. For an anamorphic magnification of $M = 1$ (the Littrow condition), the maximum angular field size in the spatial direction is:
  \begin{equation}
 \Delta \phi_{\rm y} = {}\frac{n_{\rm spatial} M{2(2N+1)}}\delta \phi_{\rm x}. \label{eqn2}
 \end{equation}
This limits the maximum slit length to 40 arc sec. on the sky. The angular field size in the spectral direction is then $\Delta \phi_{\rm x}=N\delta \phi_{\rm x} =25$arc~sec., giving an aspect ratio of 1.6.

At the detector, the width of the spectra is 80~pixels. In order to provide the ability for nod-and-shuffle (see below) this spatial field is masked down to 38arc~sec. to provide gaps between spectra of 4 pixels, thus avoiding interference or overlap between the object spectra and the sky background spectra.

\subsection{Gratings \& Resolution}
The design avoids the need to articulate the cameras. To ensure that the spectrograph resolution remains approximately constant for each grating, gratings operate in first order, at Littrow and at a constant angle of deviation and deliver the same fractional free spectral range,  $\Delta\lambda/\lambda$ to the detector in the high resolution mode $R=7000$. This corresponds to a velocity resolution of 45~km.s$^{-1}$. The dispersive elements are volume phase holographic (VPH) transmission gratings, individually manufactured to the requisite number of lines per mm.  VPH gratings offer the notable advantage that their peak efficiencies can be greater than 90\% when the substrate surfaces have been anti-reflection coated. In addition, each grating may be tilted somewhat in its mounting to optimise efficiency by tuning to the `superblaze' peak.

In the WiFeS instrument, three gratings are mounted on a slide drive in each arm of the spectrograph, two high dispersion  gratings with $R=7000$ and one low dispersion grating. To provide for this lower resolution option  ($R=3000$ or a velocity  resolution of 100~km.s$^{-1}$) we have coupled prisms to the two low-dispersion  VPH gratings to match the deviation of the beam to the high-resolution case. The dispersions of the prism and the VPH grating work in opposite senses,  so the VPH dispersion needed to be somewhat increased to compensate.

For any particular grating, the wavelength coverage and the format on the detector is fixed.  This ensures that the data reduction pipeline can be standardised -- essential if the instrument is to provide an efficient stream of reduced science data to the users. The ``blue" arm is required  to operate over the wavelength range 329 - 558 nm ($R=7000$) and 329 - 590 nm ($R=3000$), and the  ``red" camera in the 529 - 912 nm ($R=7000$) and 530 - 980 nm ($R=3000$).  Two dichroics are used with the four high resolution gratings to avoid the cut-over wavelength region, and so provide high dichroic efficiency at all wavelengths. A third dichroic is used with the two resolution gratings. In this case, the lower efficiency in the region of the dichroic cut is compensated by a generous overlap in vavelength coverage between the blue and the red arms. The modes of operation offered by the WiFeS instrument are  given in Table \ref{gratings}.

\begin{table*}
\caption{The WiFeS grating and dichroic set} 
\label{gratings}
\begin{center}       
\begin{tabular}{|l|l|l|l|l|l|l|} 
\hline
\hline
\rule[-1ex]{0pt}{3.5ex}  Grating & $U_{7000}$ & $B_{7000}$ & $R_{7000}$ & $I_{7000}$ & $B_{3000}$ & $R_{3000}$ \\
\hline
\rule[-1ex]{0pt}{3.5ex}  Lines/mm & 1948 & 1530 & 1210 & 937 & 708 & 398 \\
\hline
\rule[-1ex]{0pt}{3.5ex}  $\lambda_{\rm min} (\AA)$  & 3290 & 4184 & 5294 & 6832 & 3200 & 5300 \\
\hline
\rule[-1ex]{0pt}{3.5ex}  $\lambda_{\rm 0}$ (\AA)& 3850 & 4900 & 6200 & 8000 & 4680 & 7420 \\
\hline
\rule[-1ex]{0pt}{3.5ex}  $\lambda_{\rm max}$ (\AA)& 4380 & 5580 & 7060 & 9120 & 5900 & 9800 \\
\hline
\rule[-1ex]{0pt}{3.5ex}  Dichroic\# & 1 & 2 & 1 & 2 & 3 & 3 \\
\hline
\rule[-1ex]{0pt}{3.5ex}  $\lambda_{\rm cut} (\AA)$  & 4850 & 6200 & 4850 & 6200 & 5600 & 5600 \\
\hline 
\hline 
 
\end{tabular}
\end{center}
\end{table*} 
   \begin{figure}
   \begin{center}
   \begin{tabular}{c}
   \includegraphics[width=7cm]{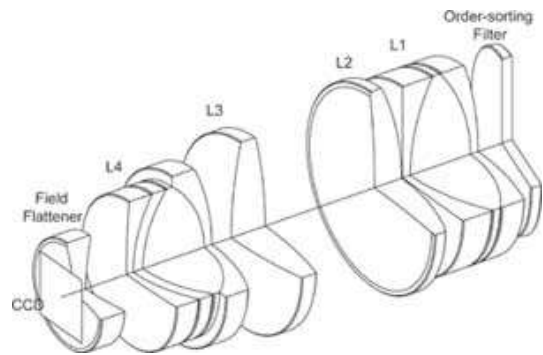}
   \end{tabular}
   \end{center}
   \caption[example] 
   { \label{fig:camera} 
  The optical configuration of the f/2.69 red camera for the WiFeS spectrograph. The blue camera configuration is very similar}
   \end{figure} 

\subsection{Camera Design}
The spectrograph presents collimated light with a beam diameter of 97 mm, to two gratings simultaneously, \emph{via} a beamsplitter. The dispersed spectra are imaged by two cameras. The ``blue" camera is required optimized for operation over the wavelength range  329 - 557nm, and the  ``red" camera in the 557 - 900nm range. Each camera has a focal length  of 261mm and each images onto a $4096\times4096$  CCD detector with $15\mu$m square pixels. An important design cost driver requirement for the camera was that all surfaces should be spherical.

The Schott 2000 and OHARA 2002 catalogues were searched for glasses with good transmission properties down to 329 nm. These were tested against each other using their differential relative partial dispersions (dRPDs). It is essential to closely match dRPDs so that secondary colour and sphero-chromatism can be minimized. Also, in some cases, it is possible to use low-powered elements  of more greatly differing dRPD to fine-tune the chromatic residuals. Two additional glasses, Calcium Fluoride and fused Silica, were also included in the glass suite because the former shows exceptionally low dispersion and the latter provides a very high transmission in the UV.

The final selection of glasses for the ``blue camera is: CaF2; SCHOTT PSK3; OHARA S-FPL51Y, BSM51Y \& S-LAL7; and fused Silica. This set delivers excellent transmission down to 329 nm with very good image quality over the entire band. Interestingly, there are far fewer well-matched glasses from which to choose for the ``red" camera.  This tends to be contrary to traditional expectations. A detailed examination of the dRPDs showed that, surprisingly, the OHARA glasses S-LAL7 and S-YGH51 were the best matches for OHARA S-FPL51Y (the core ``crown"). CaF2 becomes a progressively worse match for all these materials as one moves to longer wavelengths than 560 nm. The selections for the ÒredÓ glass set are OHARA S-FPL51Y, S-LAL7, S-YGH51 and fused Silica.  Fused Silica is used to Òfine tuneÓ the chromatic residuals in both cameras.

The cameras both use a four-component Petzval system with a thick field flattener doubling as a Dewar window. The first component is a triplet, the second and third are singlets, the fourth is a quadruplet and the field flattener singlet serving as the window to the cryostat. Neither camera is corrected for lateral colour, but this is not necessary in a spectrograph. However, the cameras are designed to correct the field distortion produced within the spectrograph. This ensures that each column on the detector corresponds to a unique spatial element -- a great advantage in data reduction. The configuration of the camera is shown in Figure \ref{fig:camera}.

  \begin{figure*}
   \begin{center}
   \includegraphics[width=12cm]{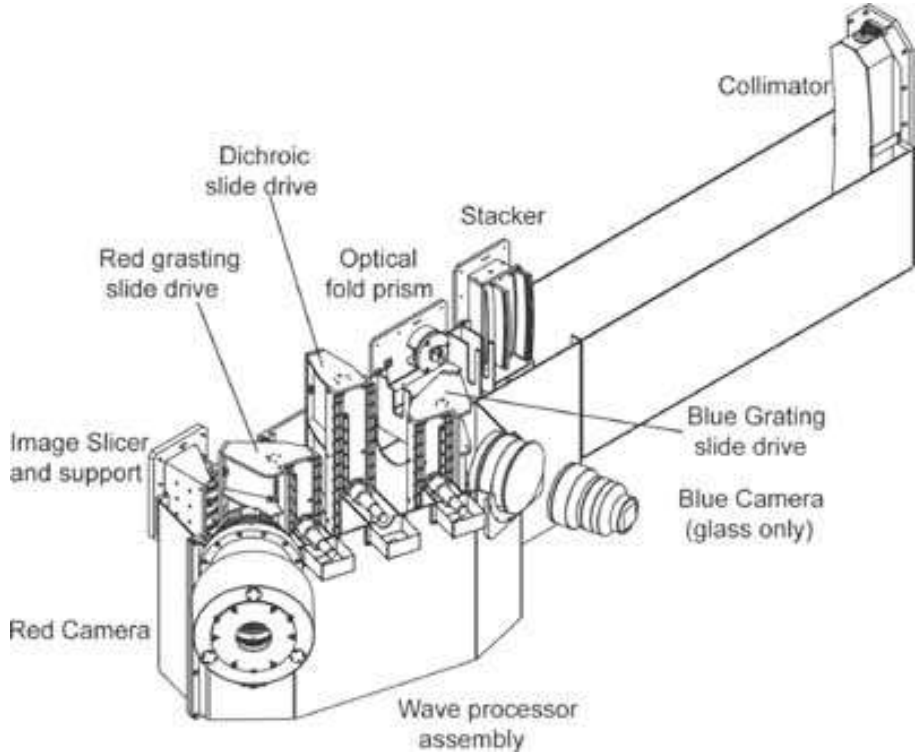}
   \end{center}
   \caption[example] 
   { \label{fig:mech} 
The overall mechanical configuration of the WiFeS spectrograph.}
   \end{figure*} 

\section{Mechanical Design}
The basic mechanical design of WiFeS is that of a fixed rigid box structure kinematically mounted on a rigid box girder platform attached to the Nasmyth B focus of the 2.3m telescope. The rotation of the field is removed by a de-rotator module mounted on the field rotator between the telescope fork and the body of the spectrograph. The calibration lamps are carried in this module. The primary function of the de-rotator is to eliminate the image rotation that occurs as the telescope tracks a star field (field spinner). It also has the secondary functions of reflecting excess field image back to the guider (focal plane unit), and reeling the electrical cables that serve itself and the guider (cable wrap).

The requirement that the spectrograph remains with fixed gravity loading is driven by the need to maintain a high spatial stability of the individual spectra on the detectors. The fixed orientation eliminates the flexure that would occur in a spectrograph free to rotate. In the case of the existing DBS, which is a spectrograph of similar size to WiFeS, this flexure leads to as much as 8 pixels of movement -- a quite unacceptable figure.

The problem with our adopted design is that the spectrograph is quite a long way from the face of the fork of the telescope. The major consideration in the design of WiFeS has been the conflicting needs for both high stiffness and low moment load. The stiffness requirement arises from the need to keep natural vibration frequencies high relative those of the telescope and so preserve telescope dynamic performance. The moment limit arises from the need to restrict eccentric loading on the azimuth bearing of the telescope and so preserve its operational life. The configuration chosen for WiFeS makes it difficult to achieve these requirements. In particular:
\begin{itemize}
\item{The spectrograph has been made stationary in orientation as a means of achieving optical stability. This has required the addition of an optical de-rotator that adds weight and complexity.}
\item{The spectrograph is displaced a long way from the fork in order to accommodate the instrument rotator, guider and de-rotator. The rotating nature of these components precludes use of an axial support system for the spectrograph, requiring the addition of a long cantilever mount. This cantilever must be slender to fit within the available space. Load moment and stiffness are then more problematic.}
\item{The collimator takes the form of an uncorrected spherical mirror in order to maximise throughput. This limits its focal ratio and makes the spectrograph unusually long and heavy. The concentric IFU also makes the collimator mirror unusually large.}
\item{ A comprehensive suite of remotely deployable beam splitter and disperser elements makes the spectrograph unusually tall and heavy.}
\end{itemize}

Careful design has resulted in a load moment of 1503 kg~m about the telescope centre, which does not significantly exceed the limit of 1500 kg m. This corresponds to an azimuth bearing lifetime of 77 years when the telescope is operating with its Caspir on Cassegrain and Imager on Nasmyth B, and 63 years for the worst-case loading of full Cassegrain and empty Nasmyth B.

Likewise, acceptable stiffness has been achieved. The lowest translational natural frequency (vertical cantilever flexure) is 59 Hz. The lowest rotational natural frequency (cantilever torsion) is 33 Hz. As required, these values are larger than the lowest natural frequency of the telescope, which is 11Hz. The associated maximum deflection due to vertical cantilever flexure is 0.072mm negligible relative to the linear image resolution in this domain of 1.75mm (this being the slit width).

Finally, the differential thermal expansion in the vertical plane over the operating temperature range of $0 - 20 ^o$C  will move the spectrograph vertically with respect to the optical axis. For a steel telescope fork and aluminium alloy spectrograph housing, the maximum shift is 0.12 mm. This also has a negligible effect on image stability.

The overall mechanical layout of the spectrograph body is shown in Figure \ref{fig:mech}. Note the kinematic truss mounting of the collimator, and the way that the vee groove mandril support for the image slicer is incorporated into the mounting. 

Note also the stacker, consisting of doublet pupil lenses, singlet field lenses, and slit image mask. This arrangement serves to capture stray light from the slicer, contributing to the excellent contrast expected in the final images. The pupil and field lenses have diameters of 9 and 10 mm respectively. All are retained in close-fitting bore holes by custom made snap rings and wave springs. With the nod-and-shuffle observing technique, the slitlet images have an angular size about the fanning axis that is slightly smaller than half the angular channel fanning pitch. This is illustrated by the separation of the perforations in the field mask. It also provides enough space between the channels to accommodate lenses that must be larger than their beam apertures.

The spherical collimator mirror has its radius of curvature centered on the fanning axis of the image slicer. It is slightly inclined so as to separate the pupil plane from the fanning plane. It is suspended on three triangular trusses. These are flexible in all deflection modes except translation in the truss plane. Together, they act as a kinematic hexapod. In addition, two brackets are supplied at mid section as safety devices that prevent the mirror from falling forward in the event of support failure. The hexapod trusses have a vertical stiffness is 40400 N/mm. For the mirror mass of 19 kg, this gives a natural vibration frequency of 1460 Hz -- very much higher than the lowest natural frequency of the telescope (11 Hz). Vibration will therefore not be excited.

\begin{figure}
   \begin{center}
   \includegraphics[width=6cm]{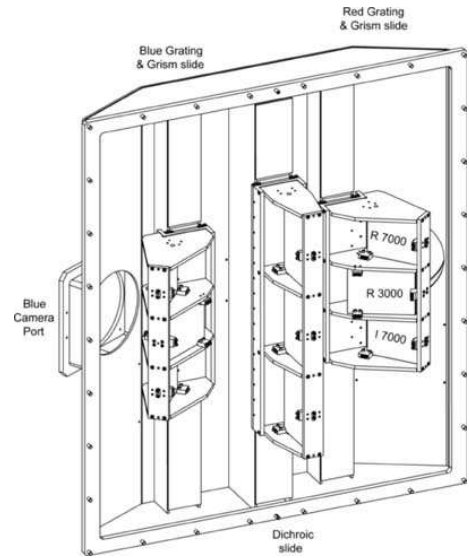}
   \end{center}
   \caption[example] 
   { \label{fig:wproc} 
The mechanical configuration of the wave processor assembly. The gratings and dichroics are permanently mounted on the slide-drives in the wave processor assembly, ensuring stability of the wavelength calibration and reproduceablity of the flux calibration. }
   \end{figure}

The details of the wave processor assembly are shown in Figure \ref{fig:wproc}. The grating and dichroics are mounted at fixed angles on slide drives which drive gratings in the direction perpendicular to the dispersion so that gratings can be rapidly exchanged with the assurance that the previous wavelength and spectrophotometric calibrations remain unchanged. Note how the low dispersion grisms are mounted in the centre of each grating drive assembly. The box plus box girder construction makes this assembly exceedingly rigid.

\begin{figure}
   \begin{center}
   \includegraphics[width=6cm]{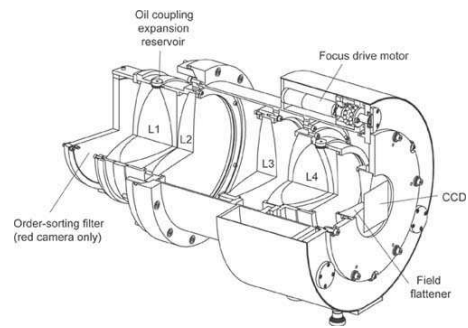}
   \end{center}
   \caption[example] 
   { \label{fig:cameramech} 
The mechanical configuration of the camera. Here the thermal stability is paramount, and the differential expansion of the glasses in the compound lenses must be accommodated by the oil coupling, as described in the text.}
   \end{figure}

The red camera is shown in Figure \ref{fig:cameramech}. Although the red and blue cameras have different optical prescriptions, they use the same mechanical design approach. The field flattener with its mounting plate is physically part of the cryostat module, where it acts as the vacuum window The mechanical structure is aluminium alloy with thin-walled cylindrical forms to achieve high stiffness and low weight. All components of multiplet lenses are oil-coupled. Circumferential o-ring seals are used to contain the oil, and volumetric compensation is provided to account for oil displacement caused by differential thermal strain, which leads to small differential changes in radii of curvature. Although these have no effect on image quality, they may lead to separation of the components, if not accommodated for.

Camera focus is accommodated with a focus stage carried on three lead screws equi-spaced on a pitch circle of 135 mm radius. Each lead screw has a pitch of 1 mm and can rotate through almost one revolution between limit switches. The travel range is therefore almost 1 mm. This is sufficient to correct temperature effects and minor errors in detector position. The three screws are independently motorized through a high reduction gearhead, and encoded. They therefore also provide tip-tilt adjustment. The drive time taken to traverse the whole range is about 30 seconds.

The detectors are mounted in the red and blue cryostats which are mechanically identical except for the vacuum window / field flattener lens. Cryocooling is provided by commercial CryoTiger closed-cycle refrigeration systems. These include remotely mounted air-cooled compressors with refrigerant lines to PT14 high performance cold heads mounted in the cryostats. These have a cooling power of 11.5 W at 94¡K.

\section{Throughput}
The need to maximize the Jaquinot advantage requires optimal throughput performance. The WiFeS  spectrograph delivers an extraordinary throughput, with peak transmission (including atmosphere, telescope, spectrograph and detector) peaking above 40\%, with $>30$\% in the wavelength range 450 - 850 nm, and $>20$\% in the wavelength range 370 - 900 nm. It exceeds the throughput of the DBS spectrograph, the spectrograph which it replaces, by typically a factor of three. This has been achieved by an aggressive effort to choose the optimum technological solution for each element throughout the optical train. The key elements of this effort will now be briefly described.

  \begin{figure}
   \begin{center}
   \includegraphics[width=8.4cm]{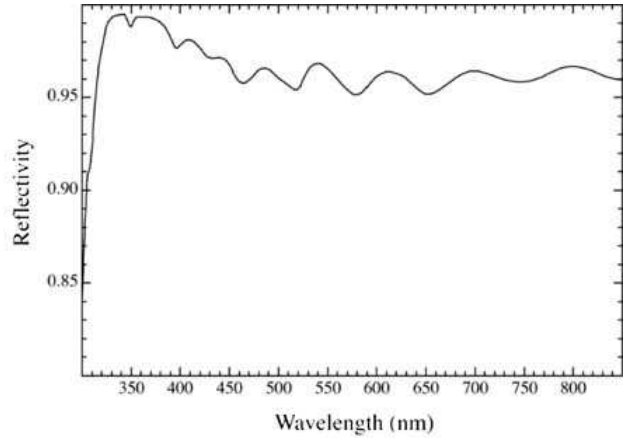}
   \end{center}
   \caption[example] 
   { \label{fig:mirrors} 
The measured reflectivity of the LLNL coatings employed for the WiFeS mirrors. At all wavelengths the reflectivity is greater than 95\%.}
   \end{figure} 

  \begin{figure}
   \begin{center}
   \includegraphics[width=8cm]{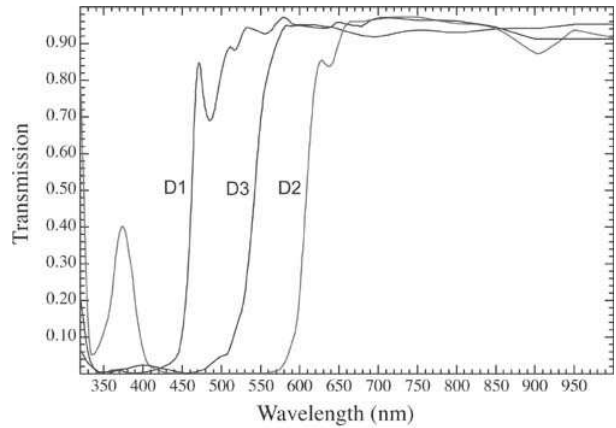}
   \end{center}
   \caption[example] 
   { \label{fig:dichroics} 
The transmission of the WiFeS dichroics. The reflectivity in the blue is assumed to be the complement of these curves.}
   \end{figure} 

 \begin{figure}
   \begin{center}
   \includegraphics[width=8cm]{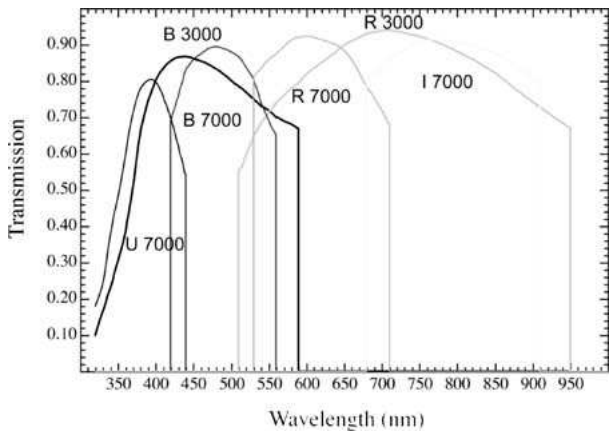}
   \end{center}
   \caption[example] 
   { \label{fig:gratings} 
The grating efficiences in their first order of the WiFeS gratings measured at their nominal operating angle of incidence. The efficiency is plotted only over the wavelength range for which each grating is intended to be used.}
   \end{figure} 

 \begin{figure}
   \begin{center}
   \includegraphics[width=8cm]{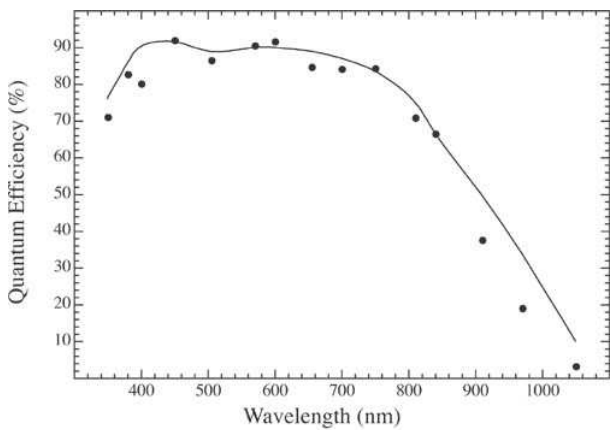}
   \end{center}
   \caption[example] 
   { \label{fig:detector} 
The measured QE of the Fairchild (blue QE enhanced) engineering test detector for WiFeS. The points are our laboratory measurements, which have an uncertainty comparable to the size of the spot, and the smooth curve shows the QE reported by the manufacturer. Note that the measued QE points fall below this curve in the red. This is a consequence of the cooling of the detector, which is necessary in order to reduce the dark noise to acceptable levels for our purpose.}
   \end{figure} 

\subsection{Mirrors}
For WiFeS, we have employed the broad-band Lawrence Livermore protected silver coating developed by Jesse Wolfe and David Sanders at LLNL \citep{Wolfe04}. These were developed for the National Ignition Facility and have most notably been used on the Keck telescope \cite{Thomas98}. This coating is very durable, passing such tests such as boiling salt water, acid and base tests, and hydrogen sulfide atmosphere. This coating delivers an average reflectivity greater than 95\% from 300 to 2500nm, and a reflectivity $>$99\% below 390nm. It uses over- and under-coats of NiCrN$_x$, nickel chromium nitride, for toughness and to retard diffusion. Layers of SiO$_2$ and Nb$_2$O$_5$ can be used to increase the reflectance at selected wavelengths. The coating can be washed with water, and wiped with a cloth, but it is stripped with sodium thiosulfate. Regrettably, these coatings are no longer available - the WiFeS mirrors were the final set to be produced by the LLNL facility.

The measured reflectivity of our mirrors is shown in Figure \ref{fig:mirrors}. This coating was used for both the image slicer and the collimator, the only mirror surfaces in the spectrograph. In addition, we manufactured  a new Nasmyth tertiary for the 2.3m telescope, and had this coated with the same material, thus improving the throughput of the telescope by $3-19$\% compared with fresh aluminium. Together, these three mirrors provide $10-60$\% increase in system throughput compared with aluminium. These gains are especially valuable both in the UV, and in the $\sim 800$nm absorption waveband of aluminium.

\subsection{Dichroics}
The WiFeS dichroics, supplied by Cascade Optics, are designed to provide maximum throughput when used with the correct gratings as shown in table 1. The dichroic cut then occurs in the unobserved wavelength band. The performance of the dichroics is shown in Figure \ref{fig:dichroics}. D3 is excellent, and D1 and D2 are very good. The blue leak apparent in these dichroics should not appreciably affect performance, since it lies largely out of band of the blue grating that will be used with them.

\subsection{Gratings}
All grating blanks are made of Ohara BSL7Y glass, which provides a high transmission below 340~nm. These were provided as 7~mm thick polished and figured substrates to the VPH grating manufacturer (ATHO$^2$L of Liege), who then fabricated the grating and assembled the sandwich.  The outside surfaces were then post polished and anti-reflection coated for optimum transmission and wavefront quality. The anti-reflection coatings are optimized for the particular wavebands, which should provide reflectivities of the order 0.5\% per air-glass surface. For the low-dispersion mode, the anti-reflection coating is applied to the outside faces of the prisms, the internal face being bonded to the VPH grating sandwiches. For the prism, the beam is incident on their outside faces at large angle. We therefore expect a poorer performance for the anti-reflection coatings; about 1.0\%.

Figure \ref{fig:gratings} shows the result of laboratory tests of the gratings at the angle of incidence that they will be actually used (22$^o$ in the case of the high dispersion gratings). The measurements below 400~nm are unreliable. The UV performance is in any case, much lower than the maufacturer's estimate in the U~7000 grating, and at the short wavelength end of the B~3000 grating. They measured the zeroth order throughput only, and the grating efficiency had to be estimated by assuming all the rest of the light was placed in the first order. Our measurements show that losses into the second order are quite important for the U~7000 and (to a lesser extent) the B~7000 gratings. Furthermore, the substrate transmission appears to be lower than expected, either due to gelatin of the grating, or the bonding glue. This is being further investigated.

The peak efficiency of these gratings can be maximized by tilting the grating to place the maximum throughput at the super-blaze angle. The effect of this was tested by repeating the grating efficiency measurements at a range of angles. All these gratings have the peak of their super-blaze within 3$^o$ of the nominal operating angle. Except for the U~7000 grating the peak efficiency is not more than $2-3$\% higher. For the U~7000 grating, the peak efficiency at the nominal operating angle is shifted to shorter wavelengths than the super-blaze peak, which provides improved efficiency at the shortest wavelengths.

All VPH gratings operated in Littrow will show a significant ghost of  zeroth order undispersed light. This was described by \citet{Wynne84} and \citet{Saunders04}. It is due to light reflected from the detector, recollimated by the camera, and reflected in $Ð1^{\rm st}-$ order by the grating to form an undispersed beam which is reimaged by the camera onto the detector. This ghost will be present at low level on the WiFeS detectors. We have evaluated this from the point of view of theory, and we do not believe it will present a significant problem. However, this remains to be verified on-telescope during the functional verification testing.

\subsection{Cameras}
The cameras provide a transmission of better than 90\% over most of the wavelength range of operation,  assuming that all air-glass surfaces have been anti-reflection coated to provide a reflection of $\le 1\%$.  In the near UV the transmission is limited by the thickness and type of glass materials used. The transmission falls to 80\% at 360 nm, and to $ < 40\% $ below 330 nm which represents the effective limit of operation of the spectrograph.

\subsection{Detectors}
The detectors are to be Fairchild $4096\times4096$ devices with 15~$\mu$m pixels. Each camera is to have a detector optimized for the wavelength of operation. Since the final scientific devices have not yet been delivered, we present results for a blue-sensitive engineering device. The quantum efficiency (QE) was measured at the RSAA detector lab on a monochromator against a standard Hamamatsu photodiode. The resulting QE curve is plotted in Figure \ref{fig:detector}. Low temperature operation of the chip reduces the QE in the red, as can be seen from these data.

In order to reduce the read-out time, we will operate these devices with dual-port readout. This will provide a readout time of $\sim 30$s.

The dark current and read-out noise of these devices is excellent. Data from 3600s dark exposures made at the operating temperature of -120C, and a system gain of 0.9e$^-$~adu$^{-1}$ and 0.89e$^-$~adu$^{-1}$ on the A-side and the B-side, respectively were analysed. These gave a net dark current of 6 and 5 e$^-$Hr$^{-1}$ on the A-side and the B-side, respectively. The readout noise for these and for shorter exposures was 3.8 e$^-$Hr$^{-1}$ for both readout ports.  These figures are sufficiently low to ensure that the spectrograph will be sky-noise limited  in its $R=3000$ mode over most of the wavelengths of operation.

 \begin{figure*}
   \begin{center}
   \includegraphics[width=13cm]{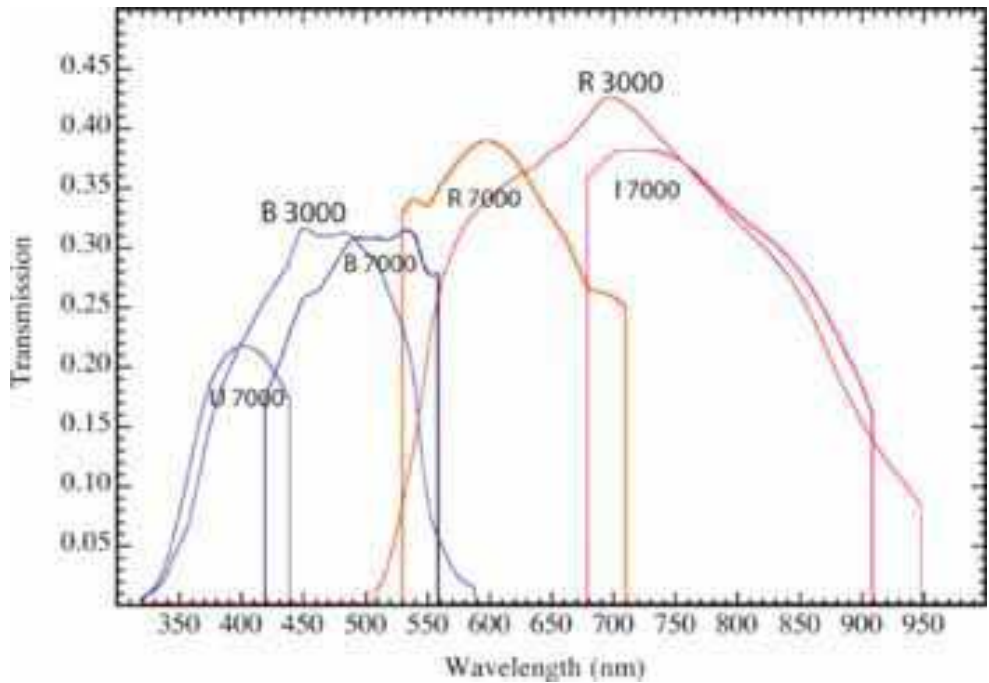}
   \end{center}
   \caption[example] 
   { \label{fig:transmission} 
The computed on-telescope end-to-end transmission of WiFeS spectrograph, including the telecope, atmosphere and detectors for both the low dispersion $R=3000$ modes and the high resolution $R=7000$ modes. Note the generous overlap between the various wavebands. This is important in ensuring accurate spectrophotometry thoughout the spectrum.}
\end{figure*} 

\subsection{System Throughput}
The end-to-end transmission expected for the WiFeS spectrograph is shown in Figure \ref{fig:transmission}. In these calculations we have included the transmission of the atmosphere, the telescope, the de-rotator optics, the slicer and stacker, the spectrograph including the collimator, camera, gratings and dichroics  and the detector performance. The primary and secondary are assumed to be uncoated Al.  For the transmissive optics, all air-glass surfaces are assumed to be A/R coated, with a reflectivity of 1\%. The bulk absorption coefficients of the various optical materials are taken from the literature.

The transmission of the instrument averages about 30\% in the 4000-9000\AA\ region, which clearly satisfies the science requirements for high throughput.
   \begin{figure*}
\begin{center}
    \includegraphics[width=15cm]{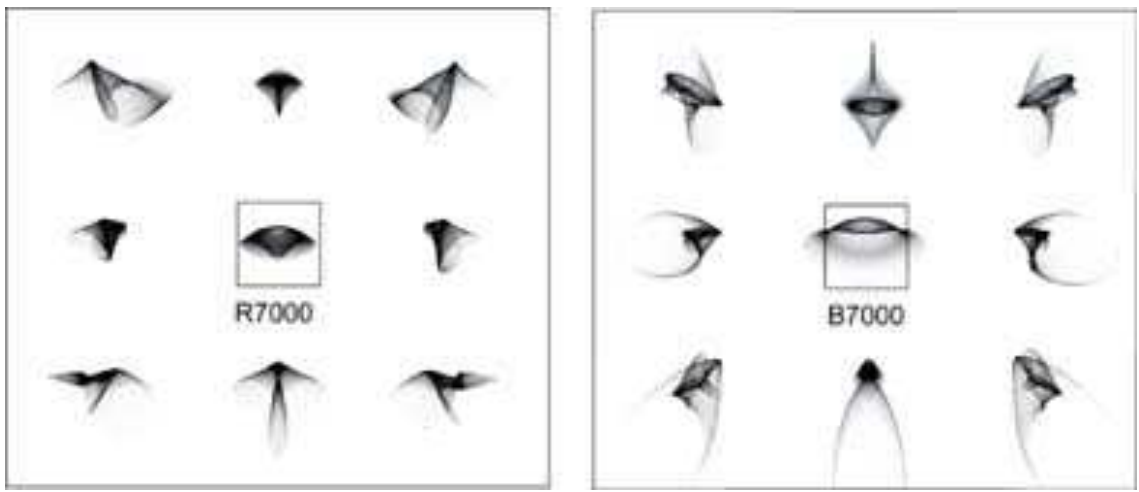}
\end{center}
     \caption[example] 
   { \label{fig:spot} 
The computed end-to-end image quality of WiFeS with the R7000 and B7000 gratings. The small box represents 2 pixels on the detector (30$\mu$m square). The nine spot diagrams for each grating correspond to positions at the centre of the science field, at each corner, and in the middle of each side. }
   \end{figure*} 

\subsection{System Sensitivity}
The effective end-to-end collecting area of WiFeS is 5000cm$^2$ (U), 15000cm$^2$ (B), 13000cm$^2$ (V), 18000cm$^2$ (R) and 5000 cm$^2$ (I). This implies a stellar photon count rate \AA$^{-1}$ at V = 15 of 4Hz (U), 21Hz (B), 13Hz (V), 11Hz (R) and 2 Hz (I). By comparison, the night sky count rate \AA$^{-1}$ at SSO in dark sky conditions is estimated to be 0.007 Hz (U), 0.017 Hz (B), 0.021 Hz (V), 0.043 Hz (R) and 0.040Hz (I). These translate to count rates per pixel on the detector at R=3000 of 0.003 Hz (U), 0.014 Hz (B), 0.019 Hz (V), 0.050 Hz (R) and 0.066 Hz (I).

WiFeS therefore becomes sky-limited at 22.0 (U), 22.7 (B), 22.0 (V), 21.1 (R), and 19.2 (I), provided that the dark current and readout noise in the detectors is below $\sim0.01$ Hz pixel$^{-1}$. This is a pessimistic estimate, since these sources of detector noise have been measured at $\sim0.003$ Hz pixel$^{-1}$ for the engineering device.

For a 22 mag star (at all wavebands), and under seeing conditions $\sim1.0$ arc sec., the exposure time needed to provide sky-limited exposures with a S/N=10 per resolution element in the $R=3000$ mode is of order 7Hr (U), 1.2Hr (B), 1.4Hr (V), 1.6Hr (R) and 4.4Hr (I). Thus, the effective limiting magnitude in the $R =3000$ mode is 22 mag, or perhaps a little fainter. For extended sources we estimate an effective limiting surface brightness of $\sim 10^{-17}$erg cm$^{-2}$ arcsec$^{-2}$ s$^{-1}$ \AA$^{-1}$ in the $R=3000$ mode.

   \begin{figure*}
\begin{center}
    \includegraphics[width=17cm]{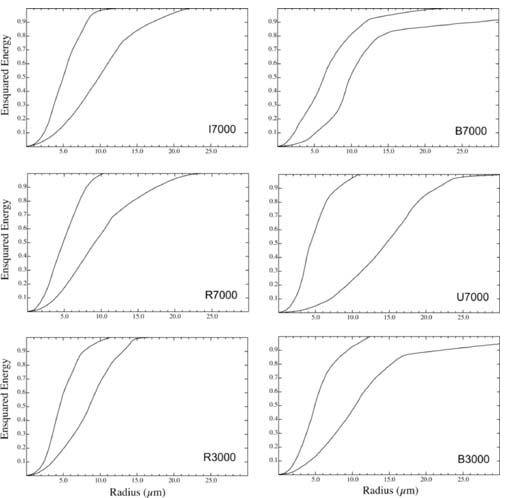}
\end{center}
     \caption[example] 
   { \label{fig:encenergy} 
The ensquared energy performance of WiFeS for each grating. The two curves are for the best and for the worst image quality within the whole of the science field. The design goal was to place at least 70\% within 15$\mu$m on the detector so that the resolution is set by the width of the entrance slit. This goal is met at all wavelengths except at the short-wavelength end of the U 7000 grating; below 350 nm where the optical design proceeded on a `best effort' basis only. }
   \end{figure*} 

\section{Image Quality}
The typical seeing expected at Siding Spring is $\sim 1.0-2.0$ arc~sec. To maximize the Felgett advantage, the instrument should be capable of using the periods of best seeing effectively, but it should not over-sample the point spread function. This requirement drives the goals of the optics design; both to match the image performance of the camera and spectrograph to the slitlet width at the detector (2 pixels or 1.0 arc~sec. on the sky) and to provide one spectral and spatial resolution element for 2 pixels on the detector.

The optical performance is limited largely by the cameras and to a lesser extent by the uncorrected collimator, which is operated slightly off-axis. The cameras are fast (F-2.6) and must operate of a wavelength range that is quite demanding. The collimator is made as fast as possible (F-12) to minimise the overall instrument length and weight. The collimator produces off-axis spherical aberration that appears as astigmatism.

The end-to-end image quality for the fore-optics, image slicer, spectrograph and camera is given in Figure \ref{fig:spot}. Here, we show both the best and the worst image quality within the field covered by the detector. In general, these correspond to a point close to the centre and a point at an extreme of wavelength near the edge of the field, but these points differ between the various gratings. 

In order to avoid degrading the spectral resolution, our objective was to ensure that the slit width is matched to 2 pixels on the detector, and that the ensquared energy is everywhere less than, or at worst equal to this projected slit width. This requires that 70\% of the encircled energy be placed within 15$\mu$m on the detector. The computed best and worst ensquared energies are shown in Figure \ref{fig:encenergy}.

Camera aberration is highly dependent on field position. Collimator aberration is independent of field position because of the concentric nature of the IFU. In some cases the collimator and camera aberrations counteract to improve image quality. It is clear that the overall image quality and hence the spectral resolution are remarkably uniform throughout the field. This is a great advantage in interpreting the reduced WiFeS data.

\section{Summary}
The WiFeS instrument is an integral field spectrograph offering unprecedented performance and field coverage. It maximizes the number of spectral elements (a total of 4096 independent elements per exposure), and the reflective image slicing design maximizes the number of spatial elements by matching spatial scale of the image at the detector to the expected seeing.  WiFeS has a science field shape (25x38 arc sec) which is well-matched to typical spatially extended science targets.  With the typical $\sim1.0$ arc sec seeing expected at Siding Spring, it presents 950 independent spatial elements.

The scientific performance of the spectrograph has been maximized through:
\begin{itemize}
\item{A reflective image slicing which ensures good spectrophotometric characteristics.}
\item{A stationary spectrograph body. This eliminates flexure, and provides a stable thermal environment, ensuring that the wavelength calibration remains fixed and stable.}
\item{Maximization of the throughput to maximize scientific productivity.}
\item{Provision of  an ``interleaved nod and shuffle'' mode. This allows photon noise-limited sky subtraction, and permits the observer to both see and to evaluate the sky-subtracted data at the telescope.}
\item{Fixed modes of operation, with all gratings and dichroics mounted within the spectrograph body. This allows rapid changes in instrumental configuration, the progressive building up of calibration libraries, and efficient batch-mode data reduction.}
\item{A design which is very well baffled against scattered light, and which has very low ghost image intensity, ensuring that fields with very large luminance contrast can be effectively observed.}
\end{itemize}

WiFeS will provide resolutions of $R=3000$ (100 km~s$^{-1}$) and $R=7000$ (45 km~s$^{-1}$) throughout the optical waveband, to a limiting magnitude of $\sim22$ for stellar sources, and $\sim 10^{-17}$erg cm$^{-2}$ arcsec$^{-2}$ s$^{-1}$ \AA$^{-1}$ for extended sources. With these performance figures, WiFeS will be highly competitive with spectrographs operating on much larger telescopes.

\begin{acknowledgments}
The authors of this paper acknowledge the receipt of the Australian Department of Science and Education (DEST) Systemic Infrastructure Initiative grant which provided the major funding for this project. They also acknowledge the receipt of an Australian Research Council (ARC) Large Equipment Infrastructure Fund (LIEF) grant LE0775546, which allowed the construction of the blue camera and associated detector module.
\end{acknowledgments}

\end{document}